\documentclass[a4paper,11pt]{article}
\pdfoutput=1 
\usepackage{comment}
\usepackage{jheppub} 
\usepackage[T1]{fontenc} 
\usepackage{xcolor}
\usepackage{subfigure}
\usepackage[export]{adjustbox}
\usepackage{amsmath}

\newcommand{\ket}[1]{\left|#1\right\rangle}
\newcommand{\bra}[1]{\left\langle #1\right|}
\newcommand{\tr}{\text{Tr}}

\newcommand{\be}{\begin{equation}}
\newcommand{\ee}{\end{equation}}

\def\text#1{{\rm #1}}

\def\avg#1{\left\langle#1\right\rangle}
\def\bra#1{\left\langle#1\right|}
\def\ket#1{\left|#1\right\rangle}

\def\abs#1{\left|#1\right|}
\def\kc#1{\left(#1\right)}
\def\kd#1{\left[#1\right]}
\def\ke#1{\left\{#1\right\}}

\def\be{\begin{equation}}       \def\ee{\end{equation}}
\def\bea{\begin{eqnarray}}      \def\eea{\end{eqnarray}}
\def\ba{\begin{array} }
\def\ea{\end{array} }

\def\nn{\nonumber}
\def\pa{\partial}
\def\=>{\Rightarrow}
\def\>{\rightarrow}

\def\vect#1{\kc{\ba{c}#1\ea}}
\def\elist#1{\left\{\ba{cc} #1\ea\right.}

\newcounter{facts}[section]

\renewcommand{\thefacts}{\arabic{facts}}

\begin{document} 

\title{\boldmath Holevo Information and Ensemble Theory of Gravity}


\author{Xiao-Liang Qi, Zhou Shangnan and Zhenbin Yang}



\abstract
{Holevo information is an upper bound for the accessible classical information of an ensemble of quantum states. In this work, we use Holevo information to investigate the ensemble theory interpretation of quantum gravity. We study the Holevo information in random tensor network states, where the random parameters are the random tensors at each vertex. Based on the results in random tensor network models, we propose a conjecture on the holographic bulk formula of the Holevo information in the gravity case. As concrete examples of holographic systems, we compute the Holevo information in the ensemble of thermal states and thermo-field double states in the Sachdev-Ye-Kitaev model. The results are consistent with our conjecture.}

\maketitle

\flushbottom

\section{Introduction}

{In recent years, various studies of low dimensional gravitational models have suggested that a simple bulk gravitational theory may be described by an ensemble of boundary theories\cite{saad2019jt,stanford2019jt,Saad:2019pqd,bousso2020gravity,marolf2020transcending,Stanford:2020wkf,Pollack:2020gfa,Afkhami-Jeddi:2020ezh,Maloney:2020nni,Belin:2020hea,Cotler:2020ugk,Chen:2020ojn, Hsin:2020mfa,Belin:2020jxr, Milekhin:2021lmq}.
More precisely, the gravitational path integral including topology changes (Euclidean wormholes) is proposed to be the holographic dual of an ensemble average of a family of quantum many-body systems, rather than a single one.\footnote{Historically, the connection between wormholes and ensemble average has been speculated by Coleman \cite{coleman1988black} and been sharpened in AdS/CFT by Maldacena-Maoz \cite{maldacena2004wormholes}. }
From the quantum information perspective, this suggests two possible different interpretations of the bulk geometry as a quantum state.
The traditional one, as in the standard holographic dictionary\cite{maldacena1999large}, is that a given bulk geometry and a bulk matter state is dual to a definite quantum state $\rho$ on the boundary when a boundary Cauchy surface is given. 
The picture of ensemble duality suggests another possibility: 
that a given bulk geometry and a boundary Cauchy surface corresponds to an ensemble average of quantum states $\rho(J)$, where the ensemble parameter $J$ can have a probability distribution $p_J$. 
The parameter $J$ may be continuous or discrete. For concreteness, we will assume it is continuous, and normalize the probability distribution as $\int dJp_J=1$. 
These two descriptions correspond to two different sets of rules of gravitational quantizations, and various information paradox can arise if we are not careful enough about distinguishing these rules.
One example that illustrates these two different descriptions is the appearance of bra-ket wormhole in a gravitational prepared state\cite{penington2019replica, chen2021bra}: 
these are global CFT states described by summation of two Euclidean semiclassical gravitational evolutions including the Hartle-Hawking geometry and a bra-ket wormhole geometry.  
On the one hand, with the standard holographic dictionary, the boundary state is a pure state with vanishing entropy, which can be correctly calculated by the island rules \cite{Almheiri:2019hni}. 
On the other hand, canonical quantization of the bra-ket wormhole geometry predicts a mixed state of the CFT. 
Such a state can be thought as an averaged description of an ensemble of different pure states depending on some unknown parameters $J$: $\rho_{a}=\int dJp_J\rho(J)$.
While these two interpretations are clear in this case, generically, the meaning of ensemble averaged states in AdS/CFT is less explored. 
}

Assuming that simple gravity is an ensemble theory, a natural question is the ``size" of the ensemble. There could be infinite possible values of parameter $J$. For example in the Sachdev-Ye-Kitaev (SYK)  model\cite{sachdev1993gapless,kitaev2014hidden,kitaev2015simple}, the natural random parameter labeling the states is the random coupling which is a set of independent Gaussian random variables. (We will discuss this model in more detail later.) However, the size of $J$ parameter space is not a physically meaningful measure of the size of the ensemble. For example, one can consider an ensemble in which $\rho(J)$ only depends on certain linear superposition of different $J$ parameters. An intrinsic measure of the size of the ensemble is the information contained in the ensemble, which can be measured by the Holevo information\cite{holevo1973bounds}. (We would like to note that Holevo information in quantum gravity has been studied for a different kind of ensemble\cite{bao2017distinguishability,bao2021holevo}.)

Consider an observer who carries quantum measurements on $\rho(J)$, and tries to learn about the random parameter $J$. In general, we can consider a projected operator valued measurement (POVM), which is defined by a set of positive Hermitian operators $M_a$ satisfying $\sum_aM_a=\mathbb{I}$. The measurement leads to an output $a$ with probability $P(a|J)={\rm tr}\kc{M_a\rho(J)}$, which corresponds to the joint probability $P(a,J)=p(J)P(a|J)$. The mutual information $I(a:J)$ between the two random variables $a,J$ measures how much information one can learn about $J$ from the measurement result $a$. For example, if $J$ is a function of $a$, the mutual information is maximal. We can define a quantum mutual information that bounds $I(a:J)$ from above. By introducing an ancilla state $\ket{J}$ which are orthogonal for different $J$, we can define the auxiliary state
\begin{align}
    \rho_{SW}=\int dJp_J\rho(J)\otimes \ket{J}\bra{J}
\end{align}
where we denote $S$ as the system and $W$ as the ancilla. The mutual information between $S$ and $W$ is the Holevo information, which provides an upper bound to the classical mutual information between $J$ and the measurement result $a$ (since measurements are quantum channels applied to $S$, which can only reduce the mutual information). The explicit formula of the Holevo information is
\begin{align}
    H=I(S:W)=S\kc{\int dJ p_J\rho(J)}-\int dJp_JS\kc{\rho(J)}
\end{align}

In this paper, we will study the Holevo information for the ensemble of states in quantum gravity. In Sec. \ref{sec:RTN}, we study the Holevo information for the ensemble of random tensor network states (RTN), which are toy models of quantum gravity. For RTN with large bond dimension, we show that the Holevo information is determined by the difference between the generalized entropy with trivial entanglement wedge and that with the actual entanglement wedge ({\it i.e.} the Ryu-Takayanagi (RT) formula\cite{ryu2006holographic}). In Sec. \ref{sec:gravity}, we make a conjecture on the holographic formula of Holevo information for quantum gravity (more precisely, for asymptotically anti de-Sitter (AdS) space and AdS coupled with non-gravitational bath). In particular, in an evaporating black hole after Page time, the Holevo information is given by the difference between the ``Hawking entropy" computed for the original geometry without replica wormhole, and the Page entropy which takes into account of the replica wormhole and entanglement island \cite{penington2020entanglement,almheiri2019entropy,penington2019replica,almheiri2020replica}. We also discussed a possible generalization of the proposed Holevo information in systems with multiple quantum extremal surfaces. In Sec. \ref{sec:SYK} we study the SYK model to gain further understanding on the Holevo information. The random parameters are the random coupling of the SYK model. We study two different ensembles of states, including the thermal states and the thermofield double (TFD) states. The Holevo information for the thermal states ensemble is given by the difference between maximal entropy and the thermal entropy. For the TFD states, the Holevo information is determined by the fermion equal-time correlation between the two sides. We discuss the holographic interpretation of these results and show that they are consistent with our conjectured bulk formula. Finally, we provide further discussion in Sec. \ref{sec:discussion}. In particular, we emphasize that the difference between RTN and gravity theory comes from the nonlocality of the random parameters in the latter case, and discuss how to possibly modify the RTN model to make it closer to the gravity theory. 


\section{Tensor network models}
\label{sec:RTN}

To understand the Holevo information $H$ in quantum gravity, we first consider the toy model of random tensor networks (RTN). An RTN is defined in the following way\cite{hayden2016holographic}. Define two quantum systems $B$ (bulk) and $S$ (boundary, in the case of AdS/CFT, or boundary and bath in the case of evaporating black hole\cite{dong2020effective}). Their Hilbert spaces are denoted as $\mathbb{H}_B$ and $\mathbb{H}_S$, with the total Hilbert space $\mathbb{H}_{BS}\equiv \mathbb{H}_B\otimes\mathbb{H}_S$. In addition, $\mathbb{H}_B=\otimes_x\mathbb{H}_x$ has a tensor factorization into different qudits, which correspond to vertices of the tensor network. An RTN state $\rho_s$ in $\mathbb{H}_S$ is defined by a random product state $\ket{V(J)}\equiv \otimes_x\ket{V_x(J_x)}\in\mathbb{H}_B$, and a state $\rho_P\in\mathbb{H}_{BS}$:
\begin{align}
    \rho_S(J)&=p_J^{-1}{\rm tr}_B\kc{\ket{V(J)}\bra{V(J)}\rho_P}\label{eq:RTN_rhoS}\\
    \text{with~}p_J&={\rm tr}_{BS}\kc{\ket{V(J)}\bra{V(J)}\rho_P}\label{eq:RTN_pJ}
\end{align}
Here $\ket{V_x(J_x)}$ is a random state in the site Hilbert space $\mathbb{H}_x$. 
More precisely, $\ket{V_x(J_x)}=U_x\ket{0}$ is obtained by applying a Haar random unitary $U_x$ on an arbitrary reference state $\ket{0}$. 
$J_x$ parameterizes this symmetric manifold with a uniform probability distribution. If we choose the normalization
\begin{align}
    \int dJ\ket{V(J)}\bra{V(J)}=\mathbb{I}_B\label{eq:completeness}
\end{align}
the ensemble of random states $\ket{V(J)}$ defines an ensemble of RTN $\rho_S(J)$ with probability $p_J$. Physically, we can consider the projection to $\ket{V(J)}$ as the consequence of a POVM, and the state $\rho_S(J)$ is the state after the measurement for the qubits that are not measured\cite{hayden2016holographic}. In this construction, $\rho_P$ can be a generic state, although conventionally a tensor network state refers to the situation when $\rho_P$ consists of a product of EPR pairs defined on links of a graph.

Now we investigate the Holevo information for this ensemble, for a subsystem $A\subset S$. According to Eq. (\ref{eq:completeness}), the averaged state $\overline{\rho_A}$ is easily determined:
\begin{align}
    \overline{\rho_{A}}=\int dJp_J\rho_A(J)={\rm tr}_{BC}\kc{\rho_P}
\end{align}
where $\rho_A(J)={\rm tr}_{C}\rho_S(J)$ is the density matrix of $A$ for fixed $J$ parameter.
Here $C$ is the complement of $A$ in $S$. 
We will consider the situation that $\rho_P$ consists of EPR pairs $\rho_P=\prod_{\avg{xy}}\ket{xy}\bra{xy}\otimes \rho_r$, and the bond dimension of each EPR pair is large, with $\rho_r$ entropy remains finite.
In such large bond dimension limit, the Renyi entropy $S^{(n)}(\rho(J))$ is independent of the random parameter $J$\cite{hayden2016holographic}, and is given by the RTN version of Ryu-Takayanagi formula\cite{ryu2006holographic}\footnote{More precisely, for a given small deviation $\delta$, there exists a critical bond dimension $D_c=\frac{\alpha}{\delta^2}e^{c_{2n}V}$ with $V$ the number of vertices in the bulk, and $\alpha,c_{2n}$ order-one consants. For $D\gg D_c$, the deviation of the Renyi entropy to the RT value is smaller than $\delta$ with a high probability: $\left|S^{(n)}\left(\rho_A(J)\right)-\min_{\Sigma\subset B}S^{(n)}_{\Sigma  A}\kc{\rho_P}\right|<\delta$ with probability $P(\delta)=1-\frac{D_c}{D}$. For more discussions about this, see Chapter 7 of Ref. \cite{hayden2016holographic}. }:
\begin{align}
    S^{(n)}\kc{\rho_A(J)}\simeq \min_{\Sigma\subset B}S^{(n)}_{\Sigma  A}\kc{\rho_P}\label{eq:RT_Renyi_RTN}
\end{align}
If we consider the large bond dimension limit such that the equation above holds for all integer $n$, the same formula applies to the von Neumann entropy by analytic continuation:
\begin{align}
    S\kc{\rho_A(J)}\simeq \min_{\Sigma\subset B}S_{\Sigma  A}\kc{\rho_P}\label{eq:RT_in_RTN}
\end{align}
The region $\Sigma$ that minimizes the righthand side is the entanglement wedge of $A$, which we denote as $\Sigma_A$. Therefore we get
\begin{align}
    H_A=S\kc{\overline{\rho_A}}-\int dJp_JS\kc{\rho_A(J)}\simeq S_A\kc{\rho_P}-S_{A\Sigma_A}\kc{\rho_P}\label{eq:HolevoRTN1}
\end{align}
If for all subregion $\Sigma$ of the bulk, $S_{\Sigma|A}\kc{\rho_P}\geq 0$, the minimization in Eq. (\ref{eq:RT_in_RTN}) will be given by an empty $\Sigma$, which corresponds to zero Holevo information. 

We would like to note that the right-hand side of Eq. (\ref{eq:HolevoRTN1}) is minus the conditional entropy of $\Sigma_A$ and $A$:
\begin{align}
    H_A=-S_{\Sigma_A|A}\kc{\rho_P}\label{eq:HolevoRTN2}
\end{align}
This conditional entropy must be negative in order for an entanglement wedge to exist. Interestingly, this is also the necessary and sufficient condition for quantum teleportation from $A$ to $\Sigma_A$ \cite{bennett1993teleporting, witten2020mini}. Our calculation shows that for random measurements, 
the Holevo information of the resulting ensemble (which measures the knowledge of the remaining quantum state about the classical measurement output) is equal to the coherent information. In the quantum setting, coherent information is positive if $\Sigma_A$ is more strongly correlated with $A$ than with the environment, which agrees with our intuition about entanglement wedge.  

Eq. (\ref{eq:HolevoRTN1}) and (\ref{eq:HolevoRTN2}) tells us that the information we can obtain from a subsystem $A$ about the random parameter $J$ is equal to the entropy difference between the contribution of trivial entanglement wedge and the actual entanglement wedge. 

\begin{figure}
    \centering
    \includegraphics[width=5in]{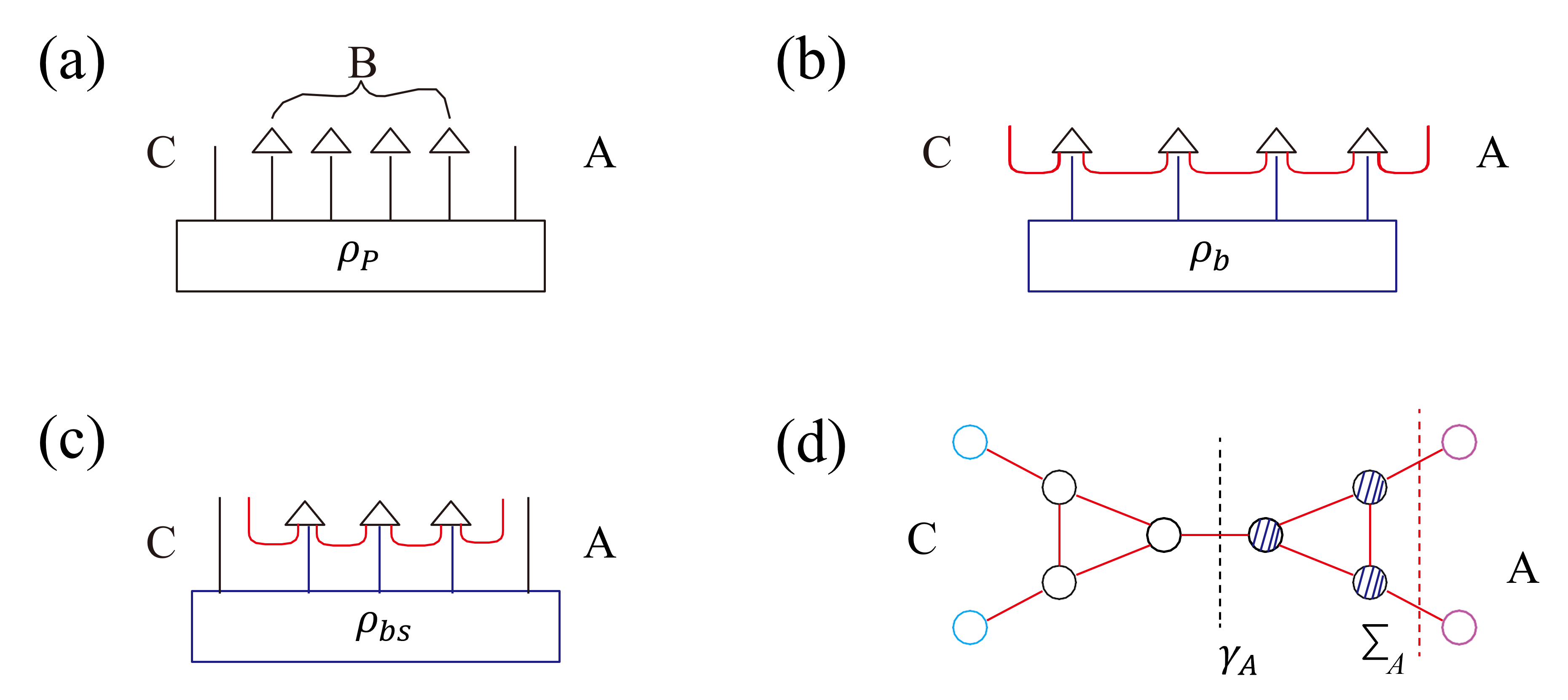}
    \caption{(a) A general random tensor network, with each triangle representing a vertex state. $\rho_P$ is generally a mixed state. Each line here (and also in subfigures (b) and (c)) corresponds to a pair of indices, labeling a complete basis in $\mathbb{H}_x\otimes\overline{\mathbb{H}}$. (b) and (c) Two examples of RTN with $\rho_P$ factorized into bulk QFT state $\rho_b$ and EPR pairs (red curves). The difference between (b) and (c) is whether the boundary is maximally entangled with its complement in $\rho_P$. (d) A more specific example of RTN, where we have only drawn the EPR pairs. In this simple case, the entropy of boundary region $A$ (purple circles on the right-hand side) is determined by the number of links at the minimal cut $\gamma_A$ (black dashed line), and the Holevo information is determined by the difference between the area of $A$ (number of links crossing the red dashed line) and that of $\gamma_A$. }
    \label{fig:RTN}
\end{figure}

To illustrate this result, we can consider two different cases, shown in Fig. \ref{fig:RTN} (b) and (c). The first example is the RTN corresponding to the AdS/CFT case, when the boundary is connected with the bulk through maximally entangled EPR pairs, and there is a bulk quantum field theory state $\ket{\Psi_b}$. (The bulk state does not have to be pure, but we write the formula for pure state case for simplicity.) The state $\rho_P=\ket{\Psi_P}\bra{\Psi_P}$ has the structure
\begin{align}
    \ket{\Psi_P}&=\otimes_{\avg{xy}\in B}\ket{xy}\otimes\ket{\Psi_b}\otimes \otimes_{x\in B,Y\in S}\ket{xY}\label{eq:RTN example1}
\end{align}
Here $\ket{xY}$ are maximally entangled EPR pairs between boundary and bulk sites. For simplicity we assume all EPR pairs have the same dimension $D$. In this case, the averaged state $\overline{\rho_A}$ is maximally mixed, and $S(\rho_A(J))$ is given by the RT formula with quantum corrections, which leads to the Holevo information
\begin{align}
    H_A&=|A|\log D-S_{\rm gen}(A)\nn\\
    &=\abs{A}\log D-\kc{\abs{\gamma_A}\log D+S_{\Sigma_A}\kc{\ket{\Psi_b}\bra{\Psi_b}}}\label{eq:H RTN case1}
\end{align}
$\gamma_A=\partial\kc{\Sigma A}$ is the minimal cut separating $A$ and the complement. This is illustrated in Fig. \ref{fig:RTN} (b) (for the general setup) and (d) (for a simple example). 


More generally, we can consider a state $\ket{\Psi_P}$ with $A$ not maximally entangled with the complement. For example, we can slightly modify the state (\ref{eq:RTN example1}):
\begin{align}
    \ket{\Psi_P}&=\otimes_{\avg{xy}\in B}\ket{xy}\otimes\ket{\Psi_{bS}}\otimes \otimes_{x\in B,Y\in S}\ket{xY}\label{eq:RTN example1}
\end{align}
as is illustrated in Fig. \ref{fig:RTN} (c). 
In this case, $A$ contains degrees of freedom from $\ket{\Psi_{bS}}$ in addition to EPR pairs, leading to the entropy
\begin{align}
    H_A=|A|\log D+S_A(\kc{\ket{\Psi_{bS}}\bra{\Psi_{bS}}})-\kc{\abs{\gamma_A}\log D+S_{A\Sigma_A }\kc{\ket{\Psi_{bS}}\bra{\Psi_{bS}}}}
\end{align}
One example of this case is the RTN model for an evaporating black hole coupled with a non-gravitational bath\cite{dong2020effective}. When $A$ is a subsystem of the bath with no direct connection to the gravitational part, there is no area law term $|A|\log D$, and the only contribution to the averaged state entropy is that of the QFT $S_A\left(\ket{\Psi_{bS}}\bra{\Psi_{bS}}\right)$. More discussions about the evaporating black hole will be carried in next section. Our discussion of the SYK model in Sec. \ref{sec:SYK} will also be related to this case.

The RT formula (\ref{eq:RT_Renyi_RTN}) for RTN comes from a replica caculation of ${\rm tr}\left(\rho_A^n\right)$ (see Appendix. \ref{app:RTN}). The two terms in Holevo information correspond to taking $n$ copies of $\rho_P$ and inserting cyclic permutation operators only in $A$ (for the first term) or in $\Sigma_AA$. This suggests that these two terms correspond to saddle points in gravitational calculation where replica wormhole is absent or present, correspondingly. We will discuss the gravity case in the next section.

\section{Conjecture about the gravity case}\label{sec:gravity}

Based on the tensor network results, we can make some conjecture on the Holevo information in AdS/CFT, or AdS/CFT coupled with bath. We start from the case of an AdS black hole coupled with a non-gravitational flat-space bath. As we learned from recent works\cite{penington2020entanglement,almheiri2019entropy,penington2019replica,almheiri2020replica}, for a large region of the radiation $A$, after Page time the entropy of $A$ is given by a nontrivial quantum extremal surface which is a boundary of an Island region $I$:
\begin{align}
    S_A=\frac{\abs{\pa I}}{4G_{N}}+S_{\rm qft}(AI)
\end{align}
If we use a replica trick and consider the calculation of Renyi entropy $S_A^{(n)}$, the bulk geometry contains a replica wormhole\cite{penington2019replica,almheiri2020replica} that connects different replicas in a region that becomes $I$ in the $n\>1$ limit. In this case, the gravitational path integral for integer $n>1$ have multiple saddle points. Besides the leading saddle point of replica wormhole, there is always a trivial saddle point which is $n$ disconnected copies of of the original geometry. Considering the tensor network results, it is natural to conjecture that after the Page time, the Holevo information is given by the difference between the trivial and nontrivial saddle points, which leads to:
\begin{align}
    H_A&=S_{\rm trivial}(A)-S_{\rm actual}(A)=S_{\rm qft}(A)-S_{\rm qft}(AI)-\frac{\abs{\pa I}}{4G_N}
\end{align}
Before the Page time, we get $H_A=0$. It should be noted that Ref. \cite{bousso2020gravity} has already proposed that the trivial saddle point corresponds to the entropy of the ensemble-averaged state.

\begin{figure}
    \centering
    \includegraphics[width=5in]{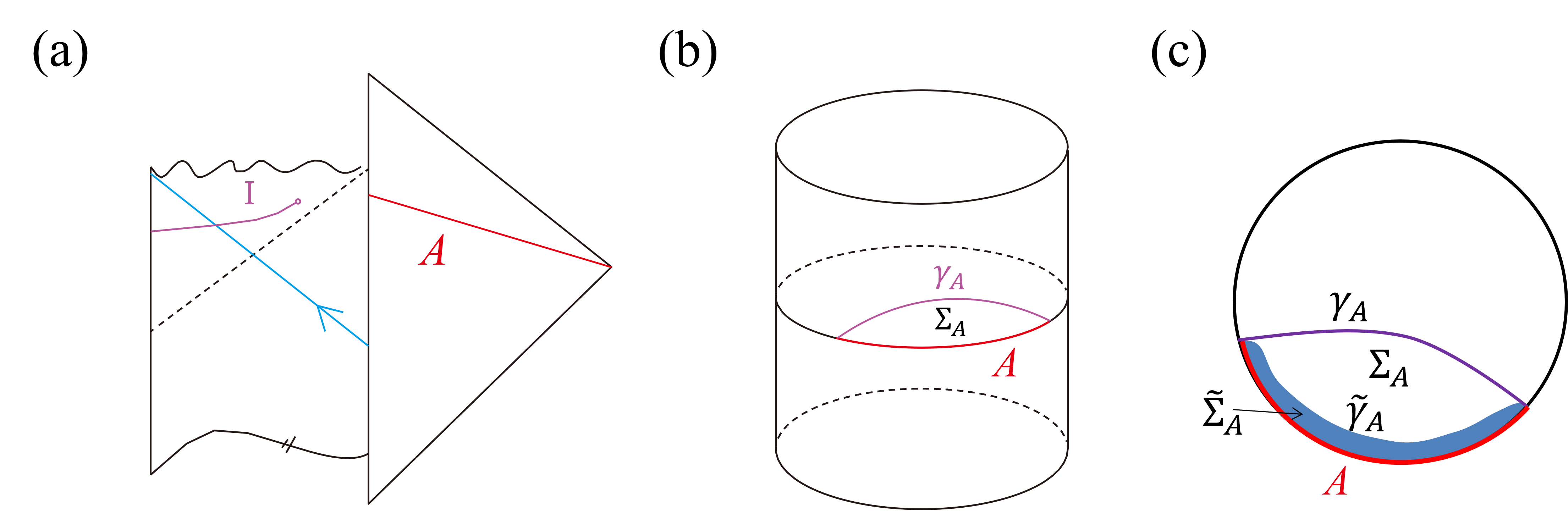}
    \caption{(a) Entanglement island in an AdS black hole coupled with flat space bath. (b) A boundary region $A$ of the AdS space, when there is no bath. (c) A top view of the spatial slice in (b), with the regularized region $\tilde{\Sigma}_A$ and its boundary $\tilde{\gamma}_A$ (see text). }
    \label{fig:AdSCFT}
\end{figure}

The situation is a bit more complicated in the AdS/CFT case without a bath. If we consider $A$ to be part of the boundary, there is no trivial saddle point in the replica calculation of $S^{(n)}(A)$ since the boundary condition already connects the different replicas. The Holevo information seems to be divergent in this case. This situation is similar to the tensor network model result in Eq. (\ref{eq:H RTN case1}). Naively one expects 
\begin{align}
    H_A=S_{\rm max}(A)-S_{\rm gen}(A)=\frac{\abs{A}-\abs{\gamma_A}}{4G_{N}}+S_{\rm qft}\left(A\right)-S_{\rm qft}(A\Sigma_A)\label{conj1}
\end{align}
Here $S_{\rm qft}(A)$ is the entropy of $A$ with trivial entanglement wedge. However, since $A$ is co-dimension-$2$ in the bulk space-time, $S_{\rm qft}(A)$ is a UV dependent quantity (just like the entropy of an infinitesimal interval in an $1+1$-dimensional CFT), which requires regularization. To regularize this quantity we introduce a region $\tilde{\Sigma}_A$ (see Fig. \ref{fig:AdSCFT} (c)), which is a small neighborhood of $A$. For example, we can cover $A$ with balls (in the boundary space) with radius $\epsilon$, and define $\tilde{\Sigma}_A$ as the union of the entanglement wedges of all the $\epsilon$-balls. (see {\it e.g.} Ref. \cite{czech2015information} for a related construction.) Then we can define
\begin{align}
    H_A=S_{gen}(A\tilde{\Sigma}_A)-S_{\rm gen}(A\Sigma_A)=\frac{\abs{\tilde{\gamma}_A}-\abs{\gamma_A}}{4G_{N}}+S_{\rm qft}\left(A\tilde{\Sigma}_A\right)-S_{\rm qft}(A\Sigma_A)\label{conj2}
\end{align}
Here $\tilde{\gamma}_A=\partial\left(A\tilde{\Sigma}_A\right)$ is a regularized version of the area of $A$, and $S_{\rm qft}\left(A\tilde{\Sigma}_A\right)$ is a regularized version of $S_{\rm qft}(A)$. 

\begin{figure}
    \centering
    \includegraphics[width=5in]{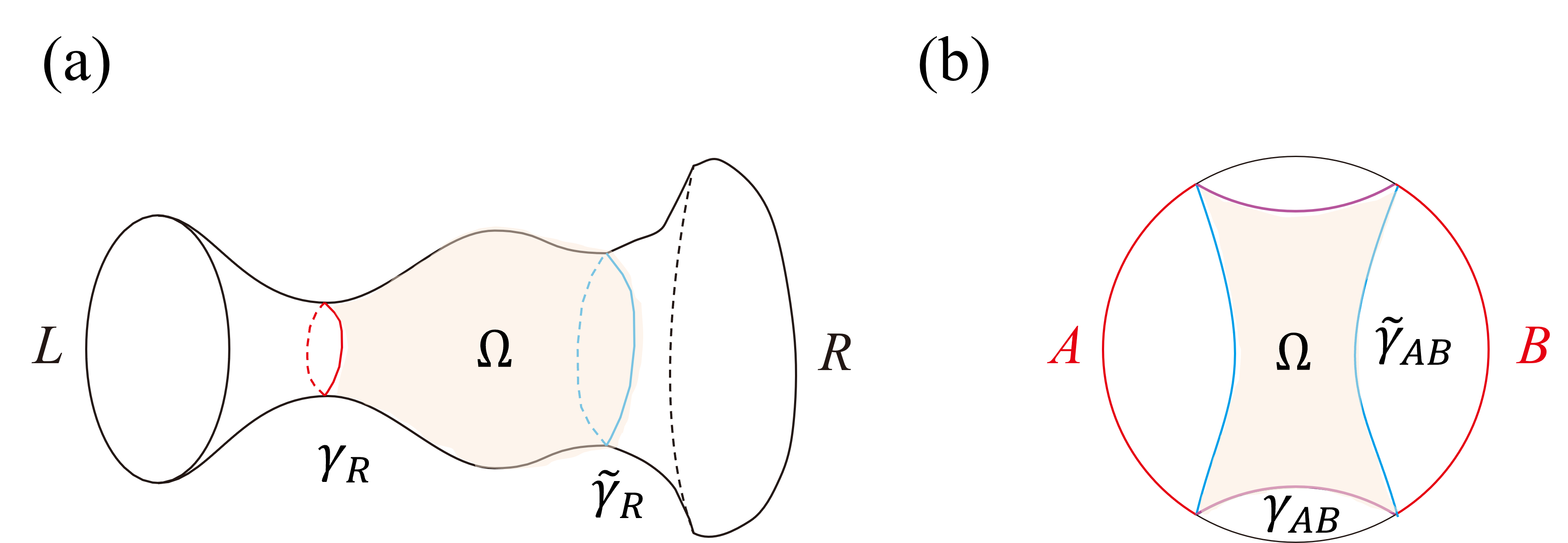}
    \caption{Two example cases with a subleading quantum extremal surface corresponding to a smaller entanglement wedge. (a) A two-sided black hole where the right boundary $R$ has an coarse-grained entropy that is larger than entanglement entropy. (b) Two regions $AB$ for which the actual entanglement wedge is connected, and there is a subleading quantum extremal surface $\tilde{\gamma}_{AB}=\gamma_A\gamma_B$ which is the disjoint HRT surface of $A$ and $B$ separately. }
    \label{fig:othersaddle}
\end{figure}

It is interesting to discuss generalizations of our conjecture to systems where there is 
a subleading saddle point corresponding to a smaller entanglement wedge. For example, one can consider a two-sided black hole corresponding to a thermofield double state $\ket{TFD}$, and then apply a unitary $U_R$ on the right boundary system. This will preserve the entanglement entropy between left and right, but if the unitary increases energy density of the $R$ system, it can lead to a bigger coarse-grained entropy, corresponding to a subleading quantum extremal surface $\tilde{\gamma}_R$ that is closer to the $R$ boundary. This is illustrated in Fig. \ref{fig:othersaddle}(a). A simpler situation is two regions in the AdS vacuum (Fig. \ref{fig:othersaddle} (b)), which has a connected entanglement wedge (when the mutual information $I(A:B)$ is nonzero to the leading order of $\frac1{G_N}$). In this case, there is a subleading saddle point corresponding to the disconnected quantum extremal surface $\tilde{\gamma}_{AB}=\gamma_A\cup \gamma_B$. This corresponds to a smaller entanglement wedge $\tilde{\Sigma}_{AB}=\Sigma_A\cup \Sigma_B\subset \Sigma_{AB}$ which is a subset of the actual entanglement wedge.

In such situation, the difference of generalized entropy contribution between these two saddles do not directly corresponds to the Holevo information we have discussed so far, since the subleading saddle point still corresponds to nontrivial connection between different replicas. However, in random tensor networks there is a Holevo information interpretation of this situation. In RTN the random parameters are locally defined at each vertex. Thus we can define a restricted ensemble by denoting $J$ as the random parameters in the tensors in a given bulk region, rather than everywhere. For example in Fig. \ref{fig:othersaddle} (a), if we define $\Omega=\Sigma_R\backslash \tilde{\Sigma}_R$ to be the region between $\gamma_R$ and $\tilde{\gamma}_R$, then averaging over the random tensor parameters in this region, denoted by $J_\Omega$, corresponds to forbidding the entanglement wedge to overlap with this region when optimizing the location of the quantum extremal surface. Thus the optimization will lead to the subleading saddle point $\tilde{\gamma}_R$:
\begin{align}
    S\kc{\int dJ_\Omega \rho\kc{J_\Omega}}=\frac{\abs{\tilde{\gamma}_R}}{4G_N}+S_{\rm qft}\kc{\tilde{\Sigma}_RR}
\end{align}
which implies that the restricted ensemble $\rho(J_\Omega)$ has the Holevo information
\begin{align}
    H_R=\frac{\abs{\tilde{\gamma}_R}-\abs{\gamma_R}}{4G_N}+S_{\rm qft}\kc{\tilde{\Sigma}_RR}-S_{\rm qft}\kc{\Sigma_RR}
\end{align}
A natural question is what the interpretation of $H_R$ is in quantum gravity. The natural candidate is an ensemble of states that share the same semi-classical geometry and are only different in the bulk region $\Omega$. For example, we can consider different bulk matter states in $\Omega$ with the same stress tensor. However, it is important to note that the ensemble we discuss is much bigger than just different bulk QFT states in $\Omega$. For example in the two-interval case in Fig. \ref{fig:othersaddle} (b), even if the system is in the AdS vacuum, such that there are no other states sharing the same stress tensor, the Holevo information is still nonzero:
\begin{align}
    H_{AB}^{(\Omega)}&=\frac{\abs{\gamma_A}+\abs{\gamma_B}-\abs{\gamma_{AB}}}{4G_N}+S_{\rm qft}\left(\Sigma_{A}\cup\Sigma_B\right)-S_{\rm qft}\left(\Sigma_{AB}\right)\nonumber\\
    &=I(A:B)-I_{\rm qft}\left(\Sigma_A:\Sigma_B\right)
\end{align}
Our interpretation is that the random parameters are not associated with different states in the bulk, but with different mappings from bulk to boundary. 
Roughly speaking, for a given semi-classical geometry, small fluctuations around this geometry defines a ``code subspace" of states which are mapped to the boundary with an isometry\cite{almheiri2015bulk}. The random parameters in the tensors physically correspond to random parameters in the bulk-to-boundary isometry. In the case of subleading saddle point, the random tensor network model suggests that we can define a two-step isometry
\begin{align}
    \Omega\longrightarrow \tilde{\gamma}_R\longrightarrow \text{boundary}
\end{align}
and $J_\Omega$ are random parameters in the first isometry. Different $J_\Omega$'s correspond to different mappings from the same bulk physics in region $\Omega$ to different boundary states.  
One should remember that the boundary Hamiltonian can correspondingly depend on $J_\Omega$, which means the boundary dynamics can appear to be $J_\Omega$ independent (at least for simple correlation functions) even if the state and the Hamiltonian depends on $J_\Omega$. We will see this more explicitly in the discussion of the SYK model in next section. 


\section{Sachdev-Ye-Kitaev model}\label{sec:SYK}

The Majorana fermion version of SYK model describes $N$ Majorana fermions with the $q$-body coupling
\begin{align}
    H_J={i^{q/2}}\sum_{j_1<j_2<...<j_q}J_{j_1j_2...j_q}\chi_{j_1}\chi_{j_2}...\chi_{j_q}
\end{align}
where $J_{j_1j_2...j_q}$'s are antisymmetric in the indices, and are independent Gaussian variables with the probability distribution
\begin{align}
    p_J&=\Omega_q^{-1}\exp\kd{-\frac1{2c_q\mathcal{J}^2}\sum_{j_1<j_2<...<j_q}J_{j_1j_2...j_q}^2}\\
    c_q&=\frac{2^{q-1}(q-1)!}{N^{q-1}q},~\Omega_q=\kc{2\pi c_q\mathcal{J}^2}^{\frac12\small{\vect{N\\q}}}
\end{align}
The SYK model has approximate conformal invariant low energy dynamics\cite{kitaev2014hidden,kitaev2015simple,maldacena2016remarks} and is proposed to approximately dual to Jackiw-Teitelboim (JT) gravity\cite{jackiw1985lower,teitelboim1983gravitation,Almheiri:2014cka,Jensen:2016pah,maldacena2016conformal,Engelsoy:2016xyb}. The random coupling $J_{j_1j_2...j_q}$ naturally defines an ensemble of states. We will study the Holevo information in this ensemble in two different scenarios.

\subsection{Thermal state}
\label{sec:SYK thermal}

The first case we consider is the thermal state 
\begin{align}
    \rho_\beta(J)=Z_J^{-1}e^{-\beta H_J}
\end{align}
In the large $N$ limit, the SYK partition function is self-averaged. More precisely,
for an order-$1$ integer $n$,
\begin{align}
    \overline{Z_J^n}\equiv\int dJp_JZ_J^{n}
\end{align}
can be written in the path integral of collective variables $G^{ab}(\tau)$ and $\Sigma^{ab}(\tau)$, with $a,b=1,2,...,n$. The path integral is dominated by a saddle point which is replica diagonal $G^{ab}(\tau)=G(\tau)\delta^{ab}$. The coupling between different replicas are suppressed by a factor $O\kc{N^{2-q}}$ \cite{kitaev2018soft}. This leads to
\begin{align}
    \overline{\log Z_J}=\log\overline{Z_J}+O\kc{N^{2-q}}
\end{align}
Thus the second term in the Holevo information $\int dJp_JS\kc{\rho_\beta (J)}$ self-averages, and is equal to the thermal entropy $S_{\rm th}\kc{\beta}$ of SYK model.

To determine the first term of the Holevo information, we need to compute
\begin{align}
    \overline{\rho}=\int dJp_J\rho_\beta(J)
\end{align}
$\overline{\rho}$ can be determined by symmetry. For each Majorana fermion operator $\chi_i$, we have
\begin{align}
    \chi_i\rho_\beta(J)\chi_i=Z_J^{-1}e^{-\beta \chi_iH_J\chi_i}
\end{align}
$\chi_i$ commutes with terms in $H_J$ that does not contain $i$, and anti-commutes with terms that contains an $i$. Denote $\tilde{J}_{j_1j_2...j_q}$ by
\begin{align}
    \tilde{J}_{j_1j_2...j_q}=\elist{-J_{j_1j_2...j_q},&~i\in\ke{j_1,j_2,...,j_q}\\
    J_{j_1j_2...j_q},&~i\notin \ke{j_1,j_2,...,j_q}}
\end{align}
Apparently $Z_{\tilde{J}}=Z_J$ and $p_{\tilde{J}}=p_J$. Thus we have
\begin{align}
    \chi_i\rho_\beta(J)\chi_i&=\rho_\beta(\tilde{J})\label{eq:Clifford symmetry}\\
    \chi_i\overline{\rho}\chi_i&=\overline{\rho}
\end{align}
In other words, $\overline{\rho}$ has to commute with all $\chi_i$, which means it has to be proportional to the identity. Thus $\overline{\rho}=2^{-N/2}\mathbb{I}$ is the maximally mixed state. Therefore we conclude that the Holevo information is given by
\begin{align}
    H_\beta=\frac N2\log 2-S_{\rm th}\kc{\beta}\label{eq:H SYK thermal}
\end{align}
This formula is consistent with the random tensor network result in Eq. (\ref{eq:H RTN case1}). The first term is the maximal entropy and the second term is the actual entropy of the state. 

In the gravity dual theory, $S_\beta$ corresponds to the RT entropy given by the dilaton value at the black hole horizon, plus the contribution of the bulk fermion field (c.f. \cite{chen2019entanglement}). As is illustrated in Fig. \ref{fig:SYK} (a), the bulk interpretation of $S\left(\overline{\rho}\right)$ is the generalized entropy of a small interval $\tilde{\Sigma}_R$, given by 
\begin{align}
    S_{gen}\left(\tilde{\Sigma}_RR\right)=\frac{\phi_\epsilon}{4G}+S_{\rm qft}\left(\tilde{\Sigma}_R\right)
\end{align}
Both terms on the right-hand side of this equation depends on the UV cutoff in the bulk, but the sum should be independent from the choice of cutoff. $S_{\rm qft}\left(\tilde{\Sigma}_R\right)$ is the entropy of a single interval for the $N$ bulk Majorana fermions. Interestingly, if we discretize the bulk AdS$_2$ fermion into a $1+1$-dimensional lattice fermion, and take the limit when $\tilde{\Sigma}_R$ is a single lattice site, the entropy $S_{\rm qft}\left(\tilde{\Sigma}_R\right)$ will reach the maximal value $\frac N2\log 2$, which suggests that $\phi_\epsilon=0$ at the lattice level. In other words, the dilaton term can be viewed as purely from integrating out bulk UV fermion modes.

\subsection{thermofield double state}\label{sec:SYK TFD}

Now we can consider a different scenario where for each $J$ we are given the thermofield double state rather than the thermal state. The thermofield double state is a purification of the thermal state, defined as follows. First we choose a maximally entangled state $\ket{I}$ between two copies of the SYK model, and then define
\begin{align}
\ket{TFD_\beta(J)}=2^{N/4}\rho_{\beta L}^{1/2}\ket{I}
\end{align}
Here $\rho_{\beta L}^{1/2}$ is the square root of the thermal density operator acting on the left system. In the limit $\beta\>0$, we get $\ket{TFD_0(J)}=\ket{I}$. It is convenient to choose $\ket{I}$ to be the state satisfying 
\begin{align}
    \kc{\chi_{iL}+i\chi_{iR}}\ket{I}=0
\end{align}
For this choice, the two-point functions satisfy
\begin{align}
    \bra{TFD_\beta(J)}i\chi_{iL}\chi_{iR}\ket{TFD_{\beta}(J)}={\rm tr}\kd{\chi_i\kc{\frac\beta 2}\chi_i(0)\rho_\beta}\equiv G\kc{\frac\beta2}
\end{align}
The average state is defined as
\begin{align}
    \overline{\rho}=\int dJp_J\ket{TFD_\beta(J)}\bra{TFD_{\beta}(J)}
\end{align}
The Holevo information is equal to the entropy of $\overline{\rho}$, since $\ket{TFD_\beta(J)}$ is a pure state and the second term vanishes. 

\begin{figure}
    \centering
    \includegraphics[width=5.5in]{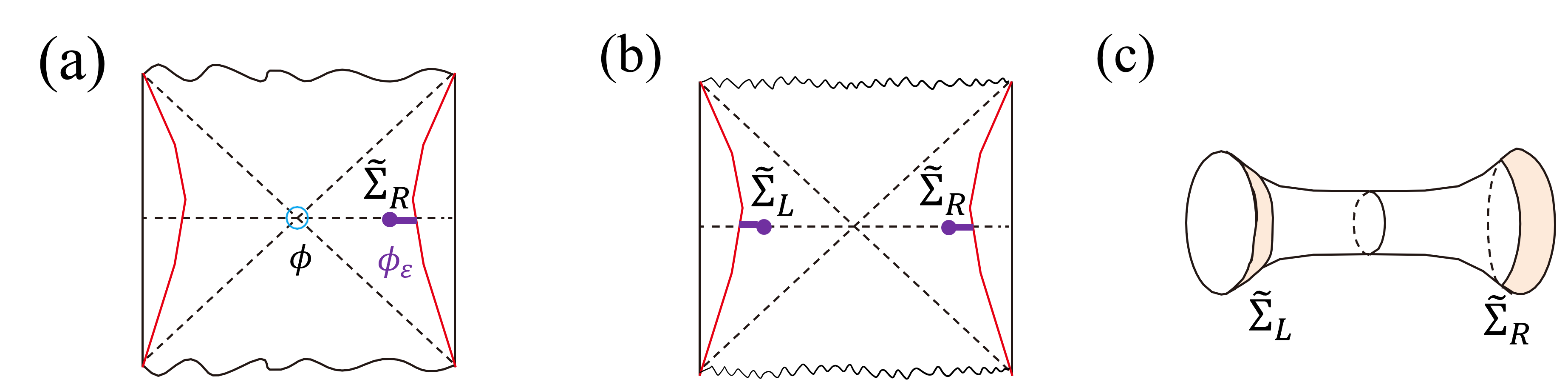}
    \caption{The holographic dual interpretation of the Holevo information, for two different ensembles of states. (a) For the ensemble of SYK thermal states, the Holevo information is the generalized entropy of a small interval $\tilde{\Sigma}_R$, minus the thermal entropy. (b) For the ensemble of thermofield double states, the Holevo information is given by the generalized entropy of two small intervals $\Sigma_L\Sigma_R$. (c) An illustration of $\Sigma_L,\Sigma_R$ in higher dimensions. }
    \label{fig:SYK}
\end{figure}
To determine $\overline{\rho}$ we notice that in the  large-$N$ limit, correlation functions of $\overline{\rho}$ satisfies the Wick theorem (which is the consequence of the saddle point approximation of SYK model). Therefore $\overline{\rho}$ is approximately a Gaussian state, which is thus determined by the two-point function
\begin{align}
    {\rm tr}\kc{\overline{\rho} i\chi_{iL}\chi_{iR}}=G\kc{\frac\beta2}
\end{align}
This condition determines
$\overline{\rho}$ to be
\begin{align}
    \overline{\rho}&=Z_a^{-1}e^{-\lambda \sum_ii\chi_{iL}\chi_{iR}},~\text{with~}\tanh\lambda=G\kc{\frac\beta 2}
\end{align}
Thus the Holevo information is 
\begin{align}
   H= S\kc{\overline{\rho}}=-N\kc{\frac{1+G\kc{\beta/2}}2\log\frac{1+G\kc{\beta/2}}2+\frac{1-G\kc{\beta/2}}2\log \frac{1-G\kc{\beta/2}}2}\label{eq:Holevo TFD}
\end{align}
There is also a more precise way to obtain this result, which we discuss in Appendix. \ref{app:SYKTFD}. 

Now we discuss more about the behavior of Holevo information (\ref{eq:Holevo TFD}) and its gravity interpretation. In the limit $\beta\>0$, $H$ vanishes because $\ket{TFD_0(J)}=\ket{I}$ is independent from $J$. In the low temperature limit $\beta\>\infty$, $H$ approaches the maximal value $N\log 2$:
\begin{align}
    H\simeq N\log 2-NG\kc{\frac\beta2}^2
\end{align}
The correction term is proportional to $\frac{N}{\kc{\beta \mathcal{J}}^{4/q}}$\cite{maldacena2016remarks}.

In the holographic dual interpretation, naively one would thought the expectation is $S\kc{\overline{\rho}}=N\log 2$ reaching the maximum, just like in the thermal state case. However, one should remember that in SYK model the bulk theory contains $N$ Majorana fermions (which is required to reproduce the $N$ boundary Majorana fermions). Therefore the correction to area law (i.e. the dilaton value) is order $N$. We can view the entropy $S(\overline{\rho})$ as the generalized entropy of the two small intervals $\Sigma_L,\Sigma_R$ in Fig. \ref{fig:SYK} (b). (For clarity, a higher dimensional picture of $\tilde{\Sigma}_L,\tilde{\Sigma}_R$ is illustrated in Fig. \ref{fig:SYK} (c).) The generalized entropy is given by
\begin{align}
    S_{gen}\kc{\tilde{\Sigma}_L\tilde{\Sigma}_R}=S_{bulk}\kc{\tilde{\Sigma}_L\tilde{\Sigma}_R}+2\phi_\epsilon
\end{align}
Here $\phi_\epsilon$ is the dilaton value at the boundary of $\tilde{\Sigma}_L$. We can write $S_{gen}\kc{\tilde{\Sigma}_L\tilde{\Sigma}_R}$ in term of mutual information:
\begin{align}
    S_{gen}\kc{\tilde{\Sigma}_L\tilde{\Sigma}_R}=S_{gen}\left(\tilde{\Sigma}_L\right)+S_{gen}\left(\tilde{\Sigma}_R\right)-I_{bulk}\left(\tilde{\Sigma}_L:\tilde{\Sigma}_R\right)
\end{align}
with $S_{gen}\left(\tilde{\Sigma}_L\right)=\phi_c+S_{bulk}\left(\tilde{\Sigma}_L\right)$ and similar for $S_{gen}\left(\tilde{\Sigma}_R\right)$. The mutual information term is only contributed by the bulk fermion. 
In the limit when the size of $\tilde{\Sigma}_L$ approaches zero, $S_{gen}(\tilde{\Sigma}_L)$ is the entropy of the averaged thermal state (Fig. \ref{fig:SYK} (a)), which should reproduce the maximal entropy $\frac N2\log 2$. The mutual information $I_{bulk}\kc{\tilde{\Sigma}_L:\tilde{\Sigma}_R}$ given by bulk fermion is expected to approach $N\log 2-H$.

It is interesting to note that at large $\beta$, the TFD state is also the ground state of the global AdS$_2$ system, which is dual to two SYK models coupled by a bilinear term\cite{maldacena2018eternal}. Therefore the Holevo information of the ensemble of TFD states can be viewed as an example of the fact that even the AdS vacuum corresponds to a nontrivial ensemble, as we discussed in Sec. \ref{sec:gravity}.


\subsection{$G-\Sigma$ action approach to the thermal state}\label{sec:SYK GSigma}

In this subsection, we will revisit the SYK thermal state discussion in Sec. (\ref{sec:SYK thermal}) with a different approach. We consider the large-$N$ effective action, with collective fields $G,\Sigma$. This approach will provide us further understanding on the gravitational saddle point that contributes to the Renyi entropy calculation of the averaged state.


The $n$-th Renyi entropy of the nonaveraged state $\rho_{\beta}(J)$ is given by the replica calculation\footnote{Here we do not include $Z_J$ in the denominator since at large N it factorizes out.}:
\begin{equation}
    \int dJ p_J\tr e^{-n \beta H_J}=\int D\Sigma D G~ \text{Pf}^N(\partial_t-\Sigma) e^{-{N\over 2}\int_{0}^{n\beta} dt_1\int_0^{n\beta} dt_2 \Sigma(t_1,t_2) G(t_1,t_2)-{J^2\over q}G^q(t_1,t_2)}.\label{eqn:G-Sigma1}
\end{equation}
Notice that since $t_{1,2}$ runs from $0$ to $n\beta$, $\Sigma(t_1,t_2)$ and $G(t_1,t_2)$ represent correlations between all the $n$ copies of the replicas. The saddle point of this path integral is just a Euclidean disk with period $n\beta$ which has nonvanishing correlation between different replica copies. 
Now let's look consider the case of the averaged state. We have:
\begin{equation}
\begin{split}
   & \int dJ_1...dJ_n p_{J_1}...p_{J_n}\tr e^{-\beta H(J_1)}...e^{-\beta H(J_n)}\\
    =&\int \prod_{i=1}^n D\Sigma_i D G_i~ \text{Pf}^N(\partial_t-\Sigma) e^{-{N\over 2}\sum_i\int_{0}^{\beta} dt_1\int_0^{\beta} dt_2 \Sigma_i(t_1,t_2) G_i(t_1,t_2)-{J^2\over q}G_i^q(t_1,t_2)}.\label{eqn:G-Sigma2}
\end{split}
\end{equation}
where $\Sigma=\text{diag}(\Sigma_1,\Sigma_2,...,\Sigma_n)$ is a block diagonal matrix of all the $\Sigma_i$s. However, $\partial_t$ term is not diagonal between replicas.:
\begin{equation}
    \text{Pf}(\partial_t-\Sigma)=\int D\psi e^{-{1\over 2}\int_0^{n\beta}dt \psi(t)\partial_t\psi(t)+{1\over 2}\sum_i\int_0^{\beta}dt^i_1\int_0^{\beta} dt^i_2\Sigma_i(t_1,t_2)\psi(t^i_1)\psi(t^i_2)};~~~t^i=t+(i-1)\beta
\end{equation}
The action (\ref{eqn:G-Sigma2}) can be viewed as a modification of the non-averaged one (\ref{eqn:G-Sigma1}), with $J^2$ in $J^2G^q(t_1,t_2)$ replaced by $J^2\theta\left(t_1,t_2\right)$. Here $\theta(t_1,t_2)=1$ or $0$ if $t_1,t_2$ belongs to the same or different replica, respectively. As a consequence, at the saddle point $\Sigma$ is always replica diagonal even if $G$ is not diagonal. 
The Schwinger-Dyson equation is:
\begin{equation}
    \begin{split}
        G(t_1,t_2)&=\left( \partial_t-\Sigma\right)^{-1};\\
        \Sigma_i(t_1,t_2)&=J^2G(t_1^i,t_2^i)^{q-1}.
    \end{split}
\end{equation}
Notice that $\left( \partial_t-\Sigma\right)^{-1}$ is a matrix with time range in $(0,n\beta)\times (0,n\beta)$ and the second equation only uses the diagonal blocks of $G$ with $t_1^i,t_2^i$ belonging to the same replica. 
It is easy to check that the solution of the SD equation is given by:
\begin{equation}
    G(t^i_1,t^i_2)=G_{\beta}(t_1^i,t_2^i),~~~G(t^i_1,t^j_2)=-2G_{\beta}(t^i_1,0)G_{\beta}(t^j_2,0),~~~i< j.
\end{equation}
where $G_{\beta}(t_1,t_2)$ is the thermal correlator.
One way to understand this result is that after average of $J$ the operator $\int dJ p(J)e^{-(\beta-t)H(J)}\chi_i e^{-t H}$ needs to be proportional to $\chi_i$ due to O(N) symmetry. The proportional constant is uniquely fixed by the thermal two point function:
\begin{equation}
    \int dJ p(J){1\over Z_J}e^{-(\beta-t)H(J)}\chi_i e^{-t H(J)}=2G_{\beta}(t,0)\chi_i.
\end{equation}
Combing this with the fact that $\int dJ p(J) {1\over Z_J}e^{-\beta H(J)}=2^{-N/2}\mathbb{I} $, one directly gets:
\begin{equation}
    \langle \chi(t_1^i)\chi(t_2^i)\rangle=G_{\beta}(t_1^i,t_2^i);~~~ \langle \chi(t_1^i)\chi(t_2^j)\rangle=-2G_{\beta}(t^i_1,0)G_{\beta}(t^j_2,0),~~~i< j.
\end{equation}

\begin{figure}
    \centering
    \includegraphics[valign = c,width=2.8in]{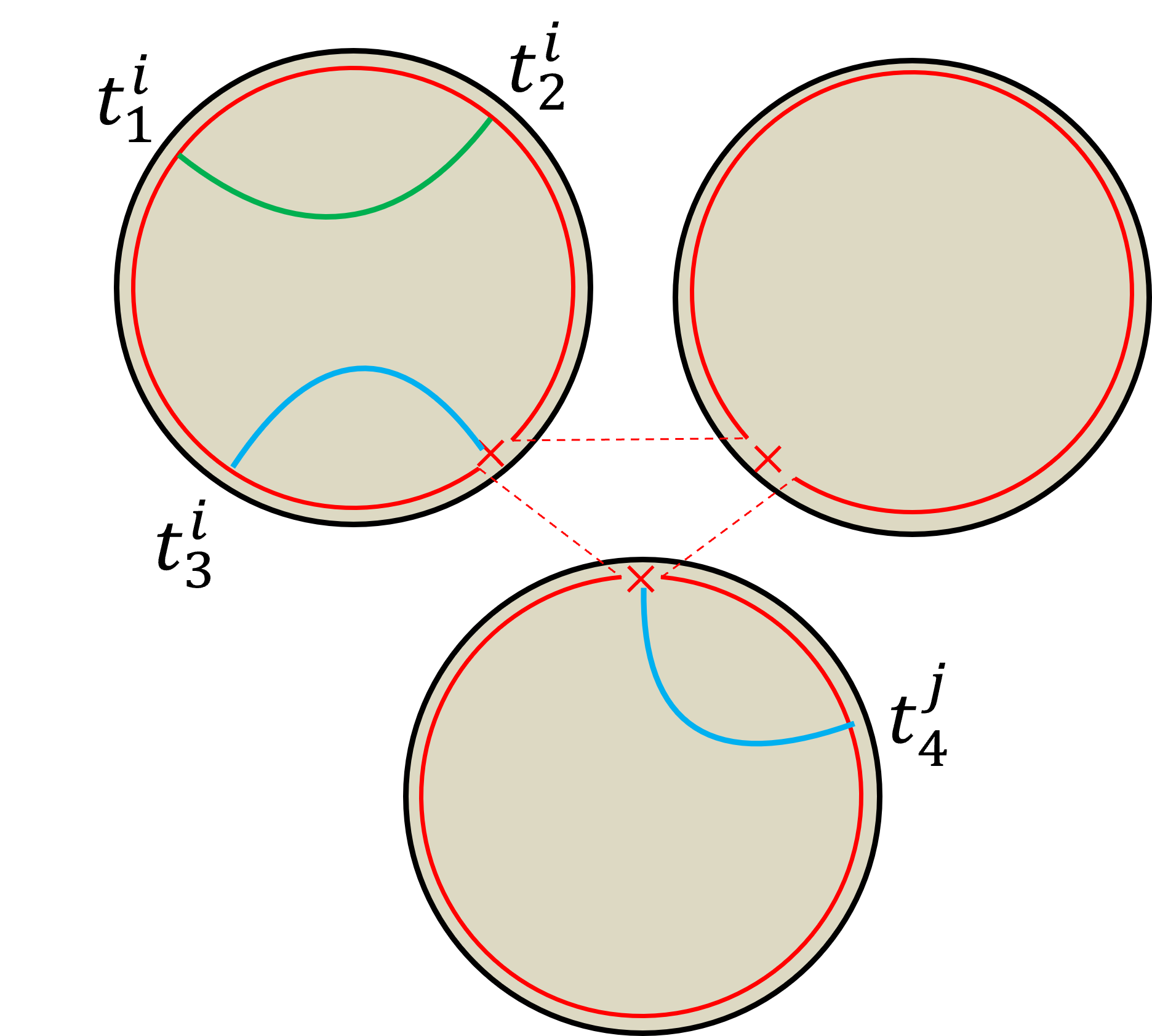}
    \caption{Illustration of the  Euclidean bulk geometry for $n$ replica of the averaged thermal state. The red circles represent the boundary of each replica with perimeter $\beta$. $n$ replica of hyperbolic disks (with $n=3$ in the figure) are glued at a branching point (red cross) at the boundary. The points connected by the red dashed lines are identified. The green and blue curves are geodesics that contribute to the same-replica two-point function $G\left(t_1^i,t_2^i\right)$ and the across-replica two-point function $G\left(t_3^i,t_4^j\right)$, respectively. 
    }
    \label{fig:flower}
\end{figure}

Taking the strong coupling limit, the above solution suggests a replica geometry in JT gravity shown in figure \ref{fig:flower}.
On such a geometry, correlators with the same replica index (the green geodesic) is the same as the thermal correlator.
Meanwhile, correlators between different replicas (the blue geodesic) will necessary factorize into product of two two-point functions since the bulk geodesic has to pass through the origin to enter a different replica system.\footnote{In pure JT such correlators will be exponential small due to the infinite distance of the origin. In SYK there is a natural IR cutoff of the boundary, which leads to a finite correlations between the off diagonal replicas.  }
Clearly, this is like freezing the disk geometry in JT gravity and only allowing the region near the boundary to glue together, which is consistent with the tensor network picture.
In the $n\rightarrow 1$ limit, the quantum extremal surface is close to the boundary which leads to the maximum entropy in the system (see Fig.\ref{fig:SYK}(a)).

To evaluate the Renyi entropy directly from $G-\Sigma$ action, we can use the following trick:
\begin{equation}
   \partial_{J^2} \tr \overline{\rho}^n=\partial_{J^2}({1\over Z^n} \int dJ_1...dJ_n p_{J_1}...p_{J_n}\tr e^{-\beta H(J_1)}...e^{-\beta H(J_n)})=0
\end{equation}
The variation of $\int dJ_1...dJ_n p_{J_1}...p_{J_n}\tr e^{-\beta H(J_1)}...e^{-\beta H(J_n)})$ with respect to $J$ is equal to ${N\over 2q}\sum_i\int\int G_i^q(t_1,t_2)$ by equation of motion.
Using the on-shell solution that $G_i(t_1,t_2)=G_{\beta}(t_1,t_2)$, this is equal to ${nN\over 2q}\int\int G_{\beta}^q(t_1,t_2)$.
This cancels exactly the variation of $Z^n$ with respect of $J$ which is also equal to ${nN\over 2q}\int\int G_{\beta}^q(t_1,t_2)$.
Therefore the Renyi entropy is independent of $J$ and we can simply evaluate the Renyi entropy in the limit of $J=0$, which gives 
\begin{equation}
    \tr \overline{\rho}^n=2^{-(n-1)N/2}.
\end{equation}
This confirms our earlier conclusion that $\overline{\rho}$ is a maximally mixed state.

The $G-\Sigma$ action is a direct analog of the Einstein Hilbert action in gravity path integral. In the language of Euclidean gravitational path integral, allowing nontrivial correlation between different replica copies corresponds to allowing short Euclidean paths between replica systems. 
We see that in the Renyi entropy calculation of the nonaveraged state, correlations between different replica systems are allowed and gives a nontrivial action.
In the Renyi entropy calculation of the ensemble averaged state, correlations between different replica systems do not contribute to the action.

The $G-\Sigma$ action approach to the averaged TFD state will be left for future work.



\section{Discussion}
\label{sec:discussion}

In this work, we explore the meaning of ensemble average in holographic system from a quantum information perspective. 
Our basic assumption is that a simple bulk geometry can be represented as an ensemble of microstates $\rho(J)$ parameterized by ensemble parameter $J$.
This assumption is true both in the random tensor network model, where the parameter $J$ is related by the local random projections in the bulk, and in the SYK model and JT gravity where $J$ is a random coupling in the Hamiltonian. 
Based on random tensor network model, we conjectured that the Holevo information of a boundary region is given by the difference between the generalized entropy of bulk regions with and without entanglement wedge (equation \ref{conj1}):
\begin{align}
    H_A=S\kc{\overline{\rho_{A}}}-\overline{S\kc{\rho_A}}=\frac{\abs{A}-\abs{\gamma_A}}{4G_{N}}+S_{\rm qft}\left(A\right)-S_{\rm qft}(\Sigma_AA).
\end{align}
Here $A$ is the boundary region, and $\Sigma_A$ is the usual entanglement wedge with quantum extremal sruface $\gamma_A$.
From the perspective of entanglement wedge recontruction, this formula suggests that all the operators in the entanglement wedge are ensemble dependent boundary operators.
Notice that even for the operators in the causal wedge, the HKLL reconstruction depends on the ensemble parameter $J$ through the dependence of the boundary Hamiltonian.
We give an explicit check of our conjecture with two SYK examples including the thermal state and the thermofield double state.

In the RTN model, the random parameters are local, associated with each bulk vertex. 
However, in SYK and JT gravity the $J$ parameter cannot be defined locally in the bulk. 
Here we would like to discuss a bit the meaning of locality of the $J$ parameters in the bulk.
In a random tensor network, the random projections $|V(J)\rangle$, after ensemble average, introduces permutations among the different bulk legs which share the same $J$ parameter.
Therefore if the $J$ parameters are local, namely they are independently random at different bulk locations, then the averaging only introduces permutations among different replica systems, without introducing any nontrivial connections within a single copy. Such permutations among replica systems can be thought of as a direct analog of replica wormholes in quantum gravity. The replica wormholes do not change simple correlation functions in a single copy of the system, but do have nontrivial effects on the bulk Hilbert space by decreasing the number of independent states.

If we modify the RTN ansatz by taking the same random state $\ket{V(J)}$ on different vertices, 
then after ensemble average there will be nontrivial permutations within a single copy of the system. 
Such permutations change the geometry of the averaged state and correspond to more general types of spacetime wormholes. 
One type of such wormhole is the one that brings a particle inside the horizon to outside at late time\cite{Saad:2019pqd}. Such wormholes are direct consequences of the random matrix behavior of the Hamiltonian, and they describe the system's memory of the initial condition. Even after ensemble averaging, such wormholes have observable consequences in a single copy system, such as the late time behavior of the two point functions.

A natural question is whether the nonlocality of $J$ parameters lead to nontrivial modification of our conjecture. In the RTN model, we expect that the additional permutations due to the nonlocality of $J$ parameters are non-perturbatively small corrections to the averaged density matrix because of the large bond dimension of the local EPR pairs in the state $|\Psi_p\rangle$.
Therefore we can safely ignore their contributions in the calculation of the Holevo information, and the conjectured formula does not change.
This is consistent with the bulk picture that even in the presence of non-local $J$ parameters, there is an approximate notion of locality due to the non-perturbative nature of the wormholes. 
What we learn from the tensor network story is that the suppression of such non-perturbative contributions stems from the entanglement structure of the parent state $\rho_P$.

We would like to conclude the discussion with a few words about the situation when such non-local effects become significant. This occurs when the wormholes proliferate. Consider a late time thermofield double state, whose interior can be approximated by a very long cylinder. When the length of the cylinder is longer than exponential of the transverse area, the wormhole will have large fluctuations by splitting and joining different locations of the cylinder. 
This completely breaks the notion of locality in the interior. 
To say it differently, the interior geometry at such late time is no longer well defined and one cannot trust the classical geometric  description.
As a toy model, consider in the RTN model a geometry with $L$ identical projections $|V(J)\rangle$ with bond dimension $D$. This can be understood as a cylinder with length $L$ and transverse area $\log D$. The average of tensor product  $\ket{V(J)}\bra{V(J)}^{\otimes L}$ results in a projection to the permutation-symmetric subspace of the $L$ bulk vertices. 
When each site has Hilbert space dimension $d$, the permutation symmetric subspace has the dimension 
\be
D_L=\binom{L+d-1}{L}\approx \begin{cases} d^L;~~~L\ll d;\\
L^{d};~~~d\ll L.
\end{cases}
\ee
In the simple model without bulk field, the vertex dimension is $d=D^2$. 
In the limit of $L\gg D^2$, the permutation symmetric subspace dimension $D_L$ is polynomial in $L$, which is much smaller than the dimension before projection. This implies that the ensemble averaging has a strong effect to single copy physical properties, although the argument is not rigorous. 
Notice that this is a truncation of the naive bulk Hilbert space in the averaged state, which is different from the truncation of the black hole Hilbert space from replica wormhole.
For instance, the effects discussed here will lead to an order-one change of simple correlation functions and the experience of an infalling observer. It will be interesting to study if there exists a new geometric description of the averaged state at late time, which is related to the typical state firewall paradox\cite{Marolf:2013dba}. 
The breakdown of the interior geometry at an exponentially late time is also expected as a consequence of the complexity-equals-to-volume conjecture\cite{Susskind:2015toa,Brown:2015bva,Brown:2015lvg}.
For a recent discussion about the behavior of the volume in JT gravity, see \cite{Yang:2018gdb,Iliesiu:2021ari}.

\section*{Acknowledgements} 

We thank Ahmed Almheiri and Henry Lin for informing us their parallel work\cite{almheiri2021}. We also noticed another recent work on a related topic\cite{renner2021black}. This work has been presented by one of us on a recent workshop\cite{qi2021geoflow}. We thank Arvin Shahbazi Moghaddam for useful discussions. This paper is supported by the National Science Foundation under grant No.\ 2111998 (XLQ and ZS), and the Simons Fundation (XLQ and ZY). This work is also supported in part by the DOE Office of Science, Office of High Energy Physics, the grant de-sc0019380.

\appendix
\section{Some more details about random tensor networks}\label{app:RTN}

In this appendix, we will discuss some more details about the replica calculation of the averaged entropy $\int dJp_JS\left(\rho_A(J)\right)$ for the ensemble of RTN given in Eq. (\ref{eq:RTN_rhoS}) and (\ref{eq:RTN_pJ}). For convenience we write these equations again here in a slightly different way:
\begin{align}
\pi_S(J)&={\rm tr}_B\kc{\ket{V(J)}\bra{V(J)}\rho_P}\label{eq:RTN_piS}\\
p_J&={\rm tr}\left(\pi_S(J)\right)\label{eq:RTN_pJapp}\\
    \rho_S(J)&=p_J^{-1}\pi_S(J)\label{eq:RTN_rhoSapp}
\end{align}
Here we have defined the unnormalized state $\pi_S(J)$. 

Using the standard replica trick
\begin{align}
    S(\rho)=-\left.\frac{\pa}{\pa n}\log{\rm tr}\left(\rho^n\right)\right|_{n\rightarrow 1}
\end{align}
we have
\begin{align}
    \int dJp_JS\left(\rho_A(J)\right)&=-\left.\int dJp_J\frac{\partial}{\partial n}\log{\rm tr}\left(\rho_A(J)^n\right)\right|_{n\rightarrow 1}\nn\\
    &=-\left.\int dJp_J\frac{\partial}{\partial n}\log{\rm tr}\left(p_J^{-1}\pi_A(J)^n\right)\right|_{n\rightarrow 1}\nn\\
    &=-\left.\int dJp_J\frac{\partial}{\partial n}\left[\log{\rm tr}\left(\pi_A(J)^n\right)-n\log p_J\right]\right|_{n\rightarrow 1}\nn\\
    &=-\left.\int dJp_J\left[\frac1{{\rm tr}\left(\pi_A(J)^n\right)}\frac{\partial{\rm tr}\left(\pi_A(J)^n\right)}{\partial n}-\frac1{p_J^n}\frac{\partial p_J^n}{\partial n}\right]\right|_{n\rightarrow 1}
\end{align}
Taking $n\rightarrow 1$ the denominator in both terms of the last line is $p_J$, so that
\begin{align}
\int dJp_JS\left(\rho_A(J)\right)&=-\left.\frac{\partial}{\partial n}\left[\int dJ{\rm tr}\left(\pi_A(J)^n\right)-\int dJ\left({\rm tr}\left(\pi_A(J)\right)\right)^n\right]\right|_{n\rightarrow 1}\label{eq:RTNpartition1}
\end{align}
Define
\begin{align}
    Z_A^{(n)}&=\int dJ{\rm tr}\left(\pi_A(J)^n\right)\\
    Z_\emptyset^{(n)}&=\int dJ\left({\rm tr}\left(\pi_A(J)\right)\right)^n=\int dJp_J^n
\end{align}
we notice that $Z_A^{(1)}=Z_\emptyset^{(1)}=1$. Thus we can further transform Eq. (\ref{eq:RTNpartition1}) into
\begin{align}
    \int dJp_JS\left(\rho_A(J)\right)&=-\left.\frac{\pa}{\pa n}\left[\log Z_A^{(n)}-\log Z_\emptyset^{(n)}\right]\right|_{n\rightarrow 1}\nn\\
    &=-\left.\frac{\pa}{\pa n}\log \frac{Z_A^{(n)}}{ Z_\emptyset^{(n)}}\right|_{n\rightarrow 1}=-\left.\frac{\pa}{\pa n}\frac{Z_A^{(n)}}{ Z_\emptyset^{(n)}}\right|_{n\rightarrow 1}
\end{align}
We prefer the last expression because $Z_A^{(n)}$ and $Z_\emptyset^{(n)}$ have the interpretation of a classical statistical model partition function, and 
\begin{align}
    -\log\frac{Z_A^{(n)}}{ Z_\emptyset^{(n)}}
\end{align}
is the free energy cost of the spin model caused by changing the boundary condition in $A$ region. The spin model comes from the fact that $\pi_S(J)$ is linear in $\ket{V(J)}\bra{V(J)}$, and 
\begin{align}
    \int dJ\ket{V(J)}\bra{V(J)}^{\otimes n}=c_n^{-1}\sum_{g\in S^n}X_g
\end{align}
with $g$ an element of the permutation group $S^n$, and $X_g$ is the permutation $g$ acting naturally by permuting different replica. $c_n$ is a normalization constant. The explicit expression of $Z_A^{(n)}$ and $Z_{\emptyset}^{(n)}$ is\cite{hayden2016holographic}
\begin{align}
    Z_A^{(n)}&=c_n^{-V}\sum_{\ke{g_x},~g_x\in S^n}{\rm tr}\kc{\rho_{AB}^{\otimes n}\otimes_{x\in B} X_{g_xx}\otimes X_{An}}\\
     Z_\emptyset^{(n)}&=c_n^{-V}\sum_{\ke{g_x},~g_x\in S^n}{\rm tr}\kc{\rho_{AB}^{\otimes n}\otimes_{x\in B} X_{g_xx}}
\end{align}

It is interesting to note that the derivation so far has not used any specific property of $\rho_P$. The averaged entropy $\int dJp_JS(\rho_A(J))$ is directly related to the analytic continuation of the free energy of the spin model. On comparison, the quantity discussed in Ref. \cite{hayden2016holographic} is the average $\int dJS^{(n)}\left(\rho_A(J)\right)$ which only corresponds to $-\frac1{n-1}\log\frac{Z_A^{(n)}}{ Z_\emptyset^{(n)}}$ in the large bond dimension limit when the numerator and denominator are self-averaging. In other words, going beyond the self-averaging limit, the averaged entropy $\int dJp_JS(\rho_A(J))$ provides a physical interpretation of the free energy cost $-\log\frac{Z_A^{(n)}}{ Z_\emptyset^{(n)}}$. 
\section{Averaged state of the thermofield double state of SYK model}\label{app:SYKTFD}

In this appendix, we carry a more detailed discussion on the averaged state of the SYK model TFD state.   It can be seen that $\overline{\rho}$ is invariant under $SO(N)$ rotation that rotate $\chi_i$ as a vector $\chi_i\rightarrow O_{ij}\chi_j$, since such rotation is equivalent to $SO(N)$ transformation of the coefficients
\begin{align}
    J_{i_1i_2...i_n}\rightarrow J_{j_1j_2...j_n}O_{j_1i_1}O_{j_2i_2}...O_{j_ni_n}
\end{align}
which preserves the probability $p_J$. The only $SO(N)$ scalar operators are
\begin{align}
    \hat{L}=\sum_i\frac{i\chi_{iL}\chi_{iR}+1}{2},~F_L=i^{N/2}\chi_{1L}\chi_{2L}...\chi_{NL},~F_R=i^{N/2}\chi_{1R}\chi_{2R}...\chi_{NR}
\end{align}
In addition, Eq. (\ref{eq:Clifford symmetry}) translates to a property of the TFD state:
\begin{align}
\chi_{iL}\chi_{iR}\ket{TFD(J)}\bra{TFD(J)}\chi_{iR}\chi_{iL}&=\ket{TFD(\tilde{J})}\bra{TFD(\tilde{J})}\\
   \Rightarrow \left[i\chi_{iL}\chi_{iR},\overline{\rho}\right]&=0
\end{align}
This symmetry condition further excludes $F_L$ and $F_R$, so that $\overline{\rho}$ is a function of $\hat{L}$. $\overline{\rho}$ can be written in the eigenstate subspaces of $\hat{L}$ as
\begin{align}
    \overline{\rho}&=\sum_{n=0}^Np_n\frac{\hat{\Pi}_n}{D_n}
\end{align}
with $\hat{\Pi}_n$ the projector to the subspace of $\hat{L}=n$, with the dimension $D_n=\vect{N\\n}$. Thus the entropy is given by
\begin{align}
    S\left(\overline{\rho}\right)=\sum_{n=0}^N\left(-p_n\log p_n+p_n\log D_n\right)\label{eq:Srhoa TFD}
\end{align}
$p_n$ can be computed using a generating function:
\begin{align}
    F(\theta)&\equiv \sum_np_ne^{in\theta}={\rm tr}\left(\overline{\rho}\hat{\Pi}_n\right)e^{in\theta}={\rm tr}\left(\overline{\rho}e^{i\hat{L}\theta}\right)\label{eq:generating function}
\end{align}
If we analytically continuate $\theta$ to $i\mu$, $F(i\mu)$ can be computed as a partition function of the coupled SYK model:
\begin{align}
    F(i\mu)=\bra{TFD(J)}e^{-\mu\hat{L}}\ket{TFD(J)}=\frac{\mathcal{Z}_\mu}{\mathcal{Z}_0}
\end{align}
which is self-averaging. $F(i\mu)$ can be computed explicitly in the large-$q$ limit of SYK model\cite{qi2019quantum}\footnote{See Eq. (5.1) of Ref. \cite{qi2019quantum}.} Without going into detail of $F(i\mu)$, we know that in the large-$N$ limit it has the form
\begin{align}
    F(i\mu)=e^{-N\mathcal{A}(\mu)}
\end{align}
with $\mathcal{A}(\mu)$ a finite function of $\mu$. Thus according to Eq. (\ref{eq:generating function}) we have
\begin{align}
    p_n=\int_0^{2\pi}\frac{d\theta}{2\pi}F(\theta)e^{-in\theta}=\int_0^{2\pi}\frac{d\theta}{2\pi}e^{-N\mathcal{A}(-i\theta)-in\theta}
\end{align}
In the large-$N$ limit, we can expand $\mathcal{A}(\mu)$ into second order of $\mu$:
\begin{align}
\mathcal{A}(\mu)&\simeq p\mu-\frac12\alpha \mu^2
\end{align}
with 
\begin{align}
    p&=\frac{\avg{\hat{L}}}{N}=\frac{1}2\left(G\left(\frac\beta2\right)+1\right)\\
    \alpha&=\frac1N\left(\avg{\hat{L}^2}-\avg{\hat{L}}^2\right)
\end{align}
This leads to 
\begin{align}
    p_n\simeq C\exp\left[-\frac{N}{2\alpha}\left(p-\frac nN\right)^2\right]
\end{align}
In the large-$N$ limit, $p_n$ is strongly peaked, which has the entropy (\ref{eq:Srhoa TFD})
\begin{align}
    S\left(\overline{\rho}\right)&\simeq \log D_{pN}+O(1)\simeq N\left[-p\log p-(1-p)\log(1-p)\right]+O(1)
\end{align}
This agrees with the Gaussian state result (\ref{eq:Holevo TFD}).

\bibliographystyle{utphys}
\bibliography{refs}

\end{document}